# How distant? An experimental analysis of students' COVID-19 exposure and physical distancing in university buildings

Bartolucci A[1,4], Templeton A[2], Bernardini G[3]

1 Institute of Security and Global Affairs (ISGA), Leiden University, The Hague, Netherlands, a.bartolucci@fgga.leidenuniv.nl
2 Department of Psychology, The University of Edinburgh, Edinburgh, Scotland, a.templeton@ed.ac.uk
3 Department of Construction, Civil Engineering and Architecture (DICEA), Università Politecnica delle Marche, Ancona, Italy, g.bernardini@staff.univpm.it
4 Corresponding author

## ABSTRACT (150)

University buildings are significant closed built environments for COVID-19 spreading. As universities prepare to re-start in-class activities, students' adherence to physical distancing requirements is a priority topic. While physical distancing in classrooms can be easily managed, the movement of students inside common spaces can pose higher risks due to individuals' proximity. This paper provides an experimental analysis of unidirectional student flow inside a case-study university building, by investigating students' movements and grouping behaviour according to physical distancing requirements.

Results show general adherence with the minimum required physical distancing guidance, but some spaces, such as corridors, pose higher exposure than doorways. Their width, in combination with group behaviour, affect the students' capacity to keep the recommended distance. Furthermore, students report higher perceived vulnerability while moving along corridors. Evidence-based results can support decision-makers in understanding individuals' exposure in universities and researchers in developing behavioural models in preparation of future outbreaks and pandemics.

## HIGHLIGHTS

- Students' COVID-19 exposure and physical distancing inside a university building is experimentally investigated, focusing on corridors and doors.
- In unidirectional flows, distances are generally compliant with safety recommendations.
- The probability of spending time at close distances is higher along the corridors, likely due to group processes.
- Questionnaire results indicate students perceive higher COVID-19 risk in corridors.
- Results provide data on COVID-19 exposure for decision-makers and simulation models.

# 1 INTRODUCTION

In response to the COVID-19 pandemic, national and local governments have adopted a variety of measures to control or reduce the spread of the virus (Capano et al., 2020; Krishnaratne et al., 2020). Such measures are mainly aimed at reducing physical contact among persons, reducing population mingling, and imposing the separation of citizens to ensure that, if they become ill or carry the virus, they will not transmit it to others (Karasmanaki & Tsantopoulos, 2021). Curfews, lockdowns, and closure of places where people gather in smaller or large numbers - including cancelling small and large gatherings (e.g., closures of shops, restaurants, shops, etc.) - occurred for an extended period. Additional measures included mandatory home working, physical distancing in public transport, and closure of educational institutions (Haug et al., 2020). As consequence, schools, including universities, cross the world have cancelled or reduced in-class activities and moved lectures and educational activities from face-to-face to online using alternative teaching activities such as online education, remote learning, or blended learning with a mix of the two (D'angelo et al., 2021). Efforts to contain COVID-19 resulted in 107 countries implementing national school closures by March 18, 2020 (Viner et al., 2020) which impacted over 87% of the world's student population (Araújo et al., 2020).

As some educational institutions and universities prepare to re-open or make a transition to blended education (in-class and online learning), administrators are currently trying to find the best solution to avoid the spread of COVID-19 to protect students, faculty, staff, and administrators. As already highlighted by (Romero et al., 2020), one of the main priorities for schools is minimizing all aspects of COVID-19 exposure. The Center for Disease Control and Prevention (CDC) created guidance for Institutions of Higher Education providing risk assessment and implementing several strategies to encourage behaviors that reduce the spread of COVID-19[1].

One of the most effective measures to prevent the spread of the disease is adopting physical distancing measures between people and reducing the number of times people come into close contact with each other (Karasmanaki & Tsantopoulos, 2021; Wismans et al., 2020; World Health Organization, 2020b).

---

[1] https://www.cdc.gov/coronavirus/2019-ncov/community/colleges-universities/considerations.html (last access: 12/4/2021)

The physical distance between two individuals can influence their movement and use of building spaces both under normal and emergency conditions. In such situations, people may wait for other people to move and coordinate as a group to stay together. Two possible reasons for close proximity among group members are the group attraction phenomena (D'Orazio et al., 2015), or having a social identity with those in the group (i.e. seeing others as fellow group members, see Alnabulsi et al., 2018). Research based in the social identity approach (Reicher et al., 2010) poses that people exhibit higher coordination with those in their group and tend to seek being close to one another. For example, group members walk more closely together with people they perceive as being in their group compared to those not in their group (Templeton et al., 2018), and will move more closely to one another to stay together when another group is present (Templeton et al., 2020). Moreover, previous research found that seeing others in the crowd as group members predicted being in a more central, denser location in the crowd (Novelli et al., 2013). Similarly, when given the choice to set up seating arrangements, people prefer to be seated closer to those they perceived to be in their group, compared with others outside their group or whose group membership they do not know (Novelli et al., 2010). However, the impact of group relations on physical distancing has not been tested during the COVID-19 pandemic when people are required to be physically distant to avoid putting others at risk. Previous literature from social psychology suggests that people are more likely to provide support to people in their group (Levine et al., 2005). In the context of COVID-19, this could mean that group members are motivated to maintain correct physical distance to keep one another safe.

Keeping physical distance can increase the length of time taken to reach a destination because of physical constraints in closed environments like corridors, stairs and at doors (Romero et al., 2020; Su et al., 2021). University buildings represent one of the most relevant scenarios in the educational context due to the potential high number of daily and contemporary users, the frequency in users' movements depending on the times of lessons and other teaching and learning activities (e.g. laboratories, libraries), as well as prolonged opening times (D'Orazio et al., 2021). Re-thinking the way students and staff use spaces in buildings and how they interact essentially means creating safety distance protocols and guidelines to be applied depending on the building features (i.e. the layout)

"that are typically designed with minimal distance" (Romero et al., 2020). Such features include corridors, as well as access to rooms and elevators.

Little information, however, is available on which measures are likely to be most effective and whether existing interventions could contain the spread of an outbreak on campus (Gressman & Peck, 2020).

Pedestrian traffic models were proposed to evaluate the effects of physical distancing-related strategies (Mohammadi et al., 2021; Romero et al., 2020), and are thus applicable to the general context of exposure models in buildings or outdoor areas that include users' behaviours and movement (D'Orazio et al., 2021; Fernandes et al., 2020; Ronchi & Lovreglio, 2020). Previous studies investigated factors affecting general physical distancing behaviours in different contexts to derive the effects of cultural, socio-economic and gender-related aspects on them (Bicalho et al., 2021; Guo, Qin, Wang, & Yang, 2021; Huynh, 2020).

Unfortunately, at the time of this research, to the authors' knowledge, experimental data on physical distances between pedestrians has mainly been limited to outdoor scenarios (Su et al., 2021). Consequently, exposure models in buildings seem to be generally based on theoretical assumptions of physical distances rather than on experimental data in indoor scenarios. Concerning educational spaces, studies exist on recommending safety measures including physical distancing, and on the consequences of physical distancing on the mental health of students (D'angelo et al., 2021; Karasmanaki & Tsantopoulos, 2021; Wismans et al., 2020) but there is still a lack of research on how students react to distancing inside the school spaces, and, mainly, in universities. Research from social psychology has found that group members in emergencies will coordinate to stay together, even potentially putting the self at risk to help others in the group (e.g., (Drury et al., 2009)), but this has only minimally been applied to non-emergency situations. Where research has focused on group coordination in corridors (e.g., Vizzari et al., 2015), or looked at social groups in classroom egress (Köster et al., 2014), this has not been at a time when physical distancing was mandatory.

This paper provides a first attempt to assess students' movement inside the university building in relation to physical distancing requirements. Strategies to ensure physical distancing in classrooms during lecture can be easily implemented because it is possible to reduce the users' density and/or move the activities in wider space by ensuring the management of the students by teaching staff, as in many workplaces (D'angelo et al., 2021; Lim et al., 2020). However, spaces connecting classrooms can be less managed and patrolled, and their use could affect students' behaviour in terms of lack of physical distancing even if this is for a limited time (D'Orazio et al., 2021), as they stay together when moving (Su et al., 2021; Templeton et al., 2020). In this sense, our study focuses on the analysis of students' movements in entering and leaving the classroom, rather than in attending lessons. This data is also compared to questionnaire-based data on individual perception of vulnerability and exposure factors while moving and using the university spaces. Analyses were carried out on a single university case study. Experimental results are also compared to previous literature and we discuss the application for behavioural simulation modelling in buildings.

## 2 MATERIALS AND METHODS

This research presents the results of a study conducted in The Hague campus of the Leiden University between October and November 2020 (semester 1, block 2) when optional activities were allowed in class within the Governmental and University regulations.

The experimental activities consisted of two main phases. In the first phase (section 2.1), we recorded free pedestrian flow of students in university spaces while entering and leaving a classroom for lessons, with the aim to assess foot traffic scenarios within the academic building in unidirectional conditions. In the second phase (section 2.2), we administered a questionnaire to investigate the students' perception of vulnerability while using the building spaces about the first step activities, and the danger perceived by them.

All the participants were informed about rules for physical distancing and voluntarily participated in the experiments. No student with mobility impairments were involved, and all the students were familiar with the building spaces for learning activities. All the students followed the University rules and regulations[2] in terms of wearing facial masks while moving inside the building and adopting personal protective measures such as physical distancing and hand hygiene.

### 2.1 TRACKING METHODOLOGIES AND TEST OF STUDENTS' FREE FLOW

The recording took place during one of the lectures on the 13$^{th}$ floor of the Stichthage building, one of the University buildings, that is located above the Central Station. The building is used for workgroups and optional in-class activities and has sufficient capacity for physical distancing. The focus was on the entrance door of the classroom (Figure 1-A) and the facing corridor to get to the room (Figure 1-B).

Students were asked to freely enter the room before the lesson and leave the room at the end of the lesson. No further instructions or barriers were provided to avoid limiting constraints in the students' flow at the door and along the corridor (e.g., controlled, or induced individuals' densities and distancing behaviours).

---

[2] https://www.staff.universiteitleiden.nl/binaries/content/assets/algemeen/reglementen/campusprotocol-v5.0-eng.pdf (last access: 12/04/2021)

*How distant? An experimental analysis of students' COVID-19 exposure and physical distancing in university buildings*
Bartolucci, Templeton, Bernardini – preprint submitted to International Journal of Disaster Risk Reduction

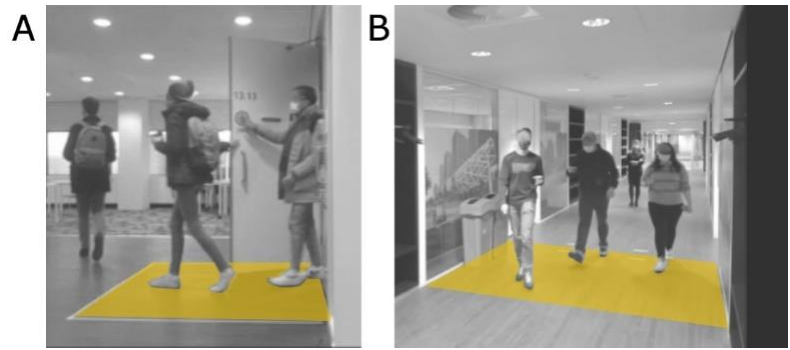

Figure 1. Two shots of the spaces where tests were performed, including the definition of the related measurement areas for motion tracking (yellow areas): A- doorway of the classroom: B- the corridor.

The entrance and egress processes were recorded using multiple cameras placed to monitor the door and the corridor, depicted in the respective views of Figure 1. Videotapes had a frame rate of 16fps. Two measurement areas (yellow areas) were defined, respectively placed: 1) inside the classroom and near to the door (Figure 1-A) to monitor the possible bottleneck effects of the door while people left the classroom; and 2) along the corridor (Figure 1-B) to evaluate unidirectional flows. The corridor-related side of the door was not investigated since preliminary analysis showed no bottlenecks because students moved into a single lane near the door (at about 1m). Either the measurement areas had a width equal to the cross-section of the element of interest (2.0 m for the overall door; 2.2 m for the corridor) and a depth of 2.0 m, so as to consider the maximum recommended physical distances from national and international health organizations and regulatory frameworks (Lim et al., 2020; Ronchi & Lovreglio, 2020; World Health Organization, 2020a). In addition, the measurement area along the corridor was placed about 2.3m far from the classroom door to avoid waiting effects due to the door usage.

The position of students inside the measurement areas were manually tracked using methodologies from previous research, with an approximation of 10cm considering the individual barycentre (Bernardini et al., 2016) and excluding individuals who moved alone and who were hidden from view by other people. The open source software Tracker for image analysis (Brown & Christian, 2011) was used to scale the scene according to in-situ measures. In particular, the tracking of students' movement at the door considered that the unidirectional motion was effectively performed into a single lane, due

*How distant? An experimental analysis of students' COVID-19 exposure and physical distancing in university buildings*
Bartolucci, Templeton, Bernardini – preprint submitted to International Journal of Disaster Risk Reductionto the door opening and crossing. The perspective filter was used for the corridor measurement area to obtain a planar view of the motion ground, where x and y coordinates of the individuals were both collected.

Using the students' positions tracking over the time, we calculated the Euclidean distances among the individuals and the deriving the time $t_{ut}$ [s] during which at least two students are placed under the following distance thresholds: (a) $t_{ut,2.0}$ for 2.0m, as the limit for close contact conditions with no masks worn and more than 15minutes of time exposure in a closed environment[3]; (b) $t_{ut,1.5}$ for 1.5m, as the recommended distance between individuals in the Netherlands[4]; and (c) $t_{ut,1.0}$ for 1.0m, that is the minimum recommended distance by World Health Organization[5]. Evaluating $t_{ut,2.0}$, $t_{ut,1.5}$ and $t_{ut,1.0}$ allow detection of the related percentage of time $pt_{ut,2.0}$, $pt_{ut,1.5}$ and $pt_{ut,1.0}$ [%] in which such conditions were reached, according to a time-exposure based approach (D'Orazio et al., 2021; Ronchi & Lovreglio, 2020). In this sense, $pt_{ut}$ values express the probability that people can stay under the considered distance thresholds while moving. The reference time was equal to the experimental tracking time. Statistical analysis of distance between individuals was also performed considering the distance range from 0 m to 2.0m. Finally, the motion speeds were retrieved and compared to results collected from previous studies related to universities and public buildings before the pandemic (Bosina & Weidmann, 2017; Fahy & Proulx, 2001).

A qualitative analyses of students' behaviours involved: (1) the number of individuals who stopped their movement to wait for other members of the same group, both while using the door and while moving along the corridor; (2) additional actions during the use of the door, such as touching the handle of any other component of the door, which could represent an additional way of virus transmission in case of contamination (Cirrincione et al., 2020).

---

[3] https://www.ecdc.europa.eu/en/covid-19/surveillance/surveillance-definitions (last access: 5/03/2021)
[4] https://www.government.nl/topics/coronavirus-covid-19/tackling-new-coronavirus-in-the-netherlands#:~:text=The%20basic%20rules%2C%20such%20as,metres%20apart%2C%20apply%20to%20everyone.&text=For%20smaller%20shops%20and%20restaurants,small%20workspaces%20or%20narrow%20corridors. (last access: 05/03/2021)
[5] https://www.who.int/emergencies/diseases/novel-coronavirus-2019/advice-for-public (last access: 05/03/2021)

Quantitative and qualitative data were assessed by separating results for the door and the corridor measurement areas. Insights on entrance and egress conditions were distinguished by considering each related experimental sample (entrance, egress, all the data).

## 2.2 QUESTIONNAIRE

Students were also asked to fill in a short complementary online survey and use 5-items Likert scales to provide further qualitative details about their perceived vulnerability to COVID-19, the danger they perceived in the environment, and to collect further qualitative experience. The questionnaire consisted of two main parts. In the first part, we asked the students a) if they were concerned about their own health due to the novel coronavirus (1 = not concerned at all; 5 = very concerned), b) how much confidence they had in the measures taken by the University to limit the spread of the novel coronavirus (1 = not confident at all; 5 = very confident) and c) how difficult was to keep physical distancing (1 = not difficult at all; 5 = very difficult) in three specific situations (walking in the corridor, entering the room and taking seat) specifying in which situation they felt most vulnerable. In the last section, we collected demographic information, such as gender and nationality, and we asked whether they have previously accessed the building during the pandemic.

## 3 RESULTS AND DISCUSSION

### 3.1 STUDENTS' FLOWS

Table 1 shows the percentage of time ($pt_{ut}$) spent at the doorway according to the thresholds (2 m, 1.5 m, and 1 m) and entrance (in) and egress (out) conditions. The results for the whole sample are also presented. For each assessed condition, the related sample dimension according to number of detected students is provided.

| % of time | DOOR | | | CORRIDOR | | |
|---|---|---|---|---|---|---|
| | in [64s; 26] | out [84s, 35] | all [148s, 61] | in [51s, 23] | out [116s, 46] | all [167s, 69] |
| $pt_{ut,2.0}$ | 21% | 25% | 23% | 49% | 54% | 53% |
| $pt_{ut,1.5}$ | 21% | 24% | 23% | 39% | 48% | 46% |
| $pt_{ut,1.0}$ | 12% | 14% | 13% | 20% | 19% | 19% |

Table 1. Percentage of time $pt_{ut}$ [%] during which at least two students are placed under the different considered distance thresholds. Door and corridor measurement areas are discussed by considering entrance (in), egress (out) and whole (all) samples. For each sample, the overall related monitored time, and the overall number of monitored students are shown in square brackets.

Filtering the distance data under the 2.0 m threshold, distances between the moving students are shown in Figure 2 according to the empirical cumulative distribution function.

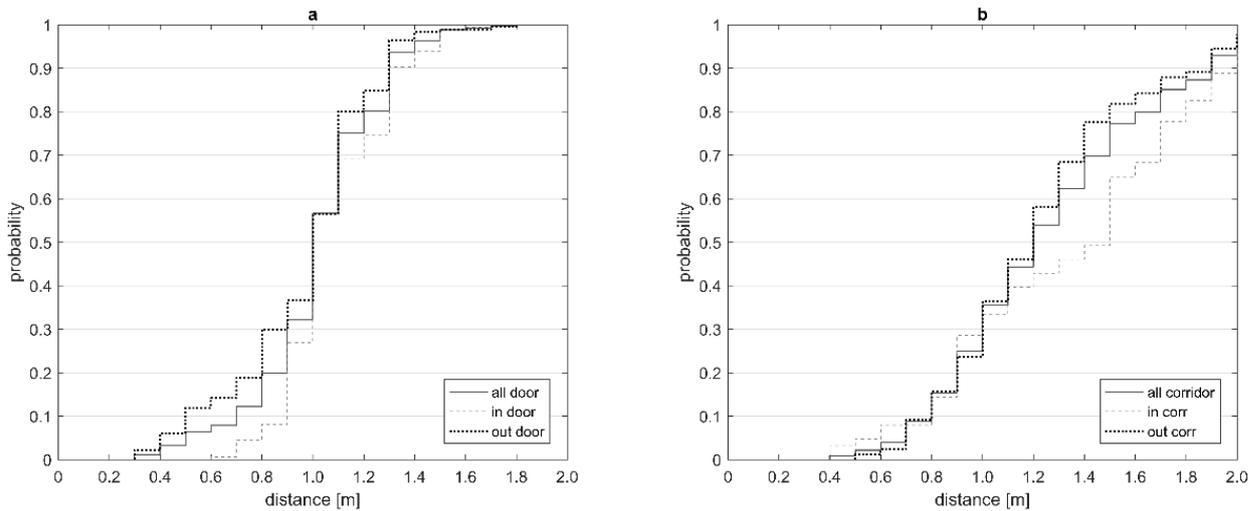

Figure 2. Empirical cumulative distribution function for door (A) and corridor (B) flows by distinguishing entrance (in – dashed lines), egress (out – dotted lines) and whole (all – continuous lines) samples.

According to Table 1, considering the measurement areas and regardless of entrance and egress conditions, students crossing the doorway seem to have a permanence time under the safety threshold lower than the one in the corridor movement. This effect could be essentially due to the single lane formation. According to Figure 2-A, the probability of students being less than the 1.0 m apart threshold is lower than the 0.4, while physical contact between the individuals are rare (distances of about 0.3 m to 0.4 m). Differences among entrance and egress samples are limited. Reasons for distances under the 2.0 m thresholds may be due to students usually standing beside the door while moving if there are other moving individuals behind them (an example is shown in Figure 5). According to this behaviour, the distance probability values are close to 1 for distance values in the range from about 1.4 m to 1.6 m (see Figure 2-A). This distance values are equal to about two people with their arms stretched out[6]. Waiting behaviours seem to be limited in the entrance sample (8%), thus affecting the minimum values of distance in these conditions as shown by Figure 2-A (0.6 m for the entrance sample rather than 0.3 m for egress sample). In contrast, such waiting behaviours were more consistent in the egress conditions (35%), because students (i.e., the first students leaving the room) seemed to prefer gathering as a group before opening the door and exiting the room. This is in line with previous research indicating how group members seek being together (e.g., staying behind to find others before leaving in emergency conditions, and leader-follower phenomena, see Drury et al., 2009; Novelli et al., 2010; Zhu, et al., 2020). An example of such behaviour is shown by Figure 3-A. While moving in the corridor area, individuals did not need to move into a single lane. Therefore, the percentage of time $pt_{ut}$ speeded by students under the safe distance thresholds is higher than in the doorway crossing, as shown by Table 1. However, Figure 2-B shows that the empirical cumulative distribution function for the corridor area has a lower slope for those at the door area, thanks to the width of the corridor allowing people to move towards higher values of physical distance, i.e., keep further apart. The distancing phenomena were more relevant when entering the room rather than when exiting the room. For instance, assuming the probability level is equal to 0.5 in both conditions, the distance with other students while entering the room was 1.6m, and while exiting the room was 1.2m. This result could be influenced by the effect of the door as a bottleneck along the path (Pradhan et al, 2020; Zhang et al., 2011). In this condition, people also avoided stopping and waiting for other

---

[6] https://multisite.eos.ncsu.edu/www-ergocenter-ncsu-edu/wp-content/uploads/sites/18/2016/06/Anthropometric-Detailed-Data-Tables.pdf (last access: 08/03/2021)

students. Distances under the safety threshold of 2.0m were limited during the time, but people stayed closer when in groups.

In comparison, while exiting the door, 26% of students who first reached the corridor (Figure 3-B1 on the right) stopped to wait for the other individuals who are still leaving the room (Figure 3-B1 on the left, the student with the red shirt), and then they began moving together when the group had gathered (Figure 3-B2).

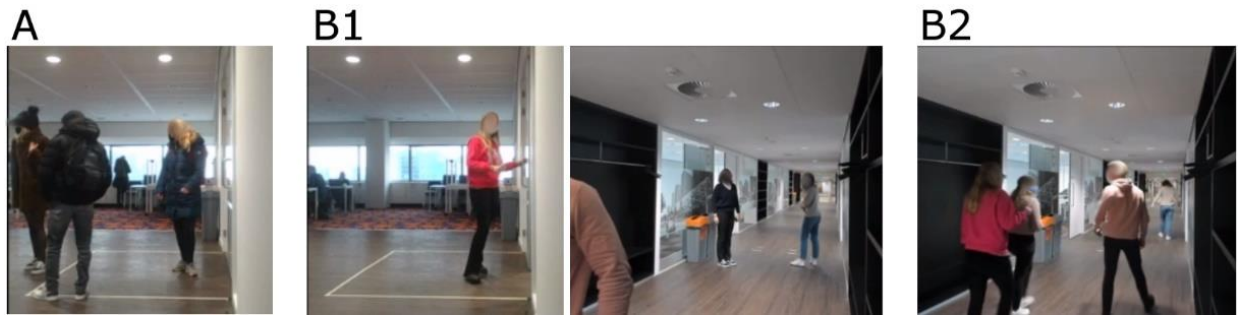

Figure 3. Typical waiting behaviours inside the room (A), from the door to the corridor (B1) and along the corridor (B2). B1 and B2 are divided by a time gap of about 6s.

Speed statistics are shown in Figure 4. The experimental median values for speed are consistently lower than values found in previous studies modelling individual motion in "normal" conditions (Lakoba et al., 2005), as well as concerning educational buildings (Bosina & Weidmann, 2017). However, the values seem to be more in line with the values for general public places (0.5 - 0.7 m/s) (D'Orazio et al., 2015; Fahy & Proulx, 2001) and research suggesting that group members reduce their speed when maintaining close proximity (Templeton et al., 2020). In this sense, the COVID-19 conditions do not seem to affect the response of the individuals, but it is noteworthy that differences between experimental outcomes and previous research may be impacted by the characteristics of the individuals, such as age, gender, grouping, clothing and physical ability (Bosina & Weidmann, 2017).

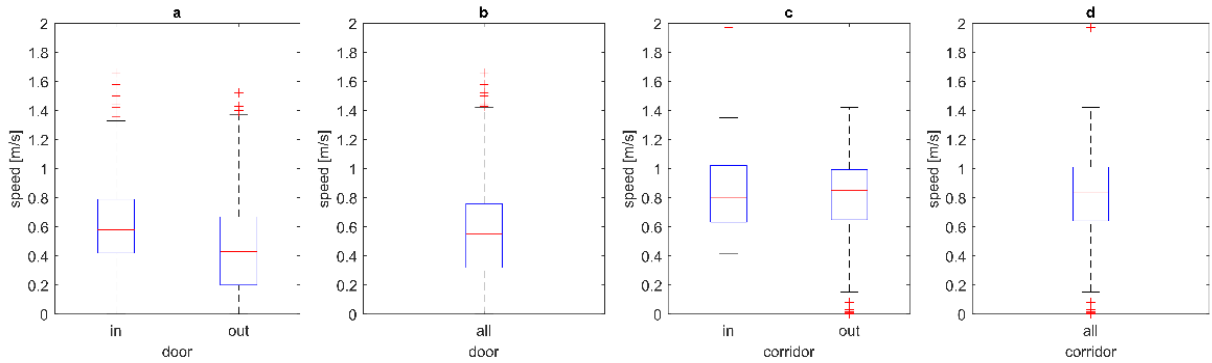

Figure 4. Boxplot representation of motion speed at the door (A, B) and the corridor (C, D) by distinguishing entrance (in), egress (out) and whole (all) samples. Outliers according to the 1.5 Inter-quartile range are shown by "+".

Finally, 85% of students touched a limited area of the door surface, as qualitatively traced by Figure 5, rather than interacting with the handle of the door.

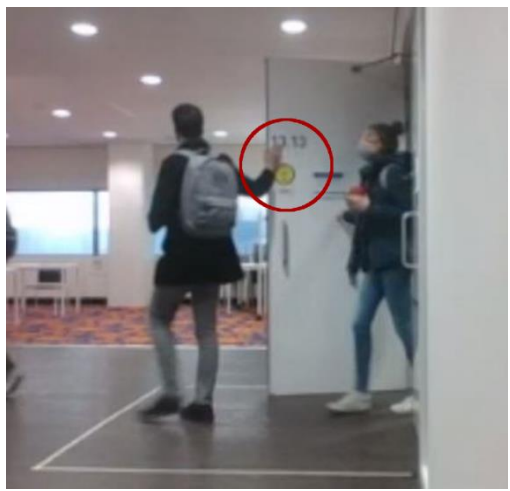

Figure 5. Qualitative behaviours of students interacting with door surface while entering the room (a frame for the videotapes): the common are for hand touch is under the red circle.

The phenomenon could be partially influenced by the fact that the door had no door lock system, thus allowing people just to push the door to enter the room. However, is it noteworthy that the same behaviours were noticed both by the first individual in the group opening the room and the following individuals. This behaviour could increase the possibility of contact-induced virus transmission if the surface is contaminated and frequently used (Cirrincione et al., 2020). Thus, regular disinfecting of all the door surfaces should be encouraged in addition to supporting correct individual hand washing. At the same time, the door pushing and holding phenomena were limitedly performed to wait for other individuals while moving, especially while entering the room (8% of students), thus highlighting how the regular distance between the individuals seems to be the leading factor influencing movement through doorways.

## 3.2 QUESTIONNAIRE

A total of 38 students took part in the survey, of those 51% were females, 86% were Dutch, and 91% had experience accessing the building. Overall, the questionnaires results suggest students were unconcerned to moderately concerned about their own health ($M = 2.82$, $SD = 1.12$). They also were confident in the measures taken by the University. Although average scores tend to underline neutral to fairly confident values ($M = 3.68$, $SD = 0.98$), 70% of the students stated they were fairly to very confident (score from 3 to 5 in the related Likert scale).

Additionally, outcomes from experimental analyses of Section 3.1 are also supported. Students rated quite low difficulty at the corridor ($M = 2.25$, $SD = 1.30$). They reported mainly having difficulty keeping physical distancing when entering the room ($M = 2.40$, $SD = 1.25$), that is while crossing through the doorway. This corresponds to the values in the cumulative distribution function in Figure 2-A, that are shifted towards lower values (e.g., close to 1m). As discussed in Section 3.1, this difficulty was mitigated by the single lane formation while crossing the door, thus limiting $pt_{ut}$ as shown by Table 1.

Finally, students seemed to be aware of the exposure while moving along the corridor, since 44% of the students felt more vulnerable in this space due to limited ability to keep the required physical distance. Participants reported higher vulnerability (i.e., scores 4 or 5 in the Likert scale) due to keeping physical distance in the corridor (44%) than at the door (38%). The wider cross section of the corridor allowed students to keep further distances from surrounding individuals, as shown by Figure 2-B values that are shifted towards 2m. However, physical grouping phenomena seemed to affect the overall exposure over the time, as shown by the $pt_{ut}$ values higher than those at the door crossing (see Table 1). The mean rating for difficulty keeping the required physical distance from others along the corridor was slightly lower than when crossing through the doorway. Although further investigation into this topic is needed, this result could be considered in line with previous research suggesting that higher the number of people in the group, higher the probability that they will move under the physical distance threshold and that some people cluster into close physical groups even during the COVID-19 pandemic (Su et al., 2021).

Finally, students reported that they did not feel particularly exposed to COVID-19 risks while taking a seat in the classroom. This result could be influenced by the safety protocols in the rooms, such as reduced numbers of people and staff support to maintain safety (D'angelo et al., 2021).

## 3.3 KEYFINDINGS AND LIMITATIONS

Results show that students generally adopted physical distancing when moving inside the university buildings. However, the presence of environmental constraints, such as doors and corridors, affected their distancing behaviour. There was lower difficulty to keep appropriate distance in the corridor due to the wider space, but the tendency to behave and walk-in groups - especially after the lecture - increased their feeling of vulnerability. The use of doors was associated with a bottleneck phenomenon that resulted in the creation of a single lane while crossing the doorway, and in reduced distance between students. The questionnaire supported the experimental results and highlighted that the perceived vulnerability was higher in the corridor, especially when leaving the room after group work, mainly because students tended to maintain proximity to their groups and exit with other people rather than individually. From a quantitative point of view, this work provides the frequency and probability of being under the required physical distance threshold. This kind of data could be useful to develop exposure assessment tools and models and hence to evaluate the experimental-based possibility of close contact between individuals using facial masks inside university common spaces (D'Orazio et al., 2021; Romero et al., 2020; Ronchi & Lovreglio, 2020).

Importantly, most research investigating group movement and clustering has used top-down assumptions of group cohesion based on observed proximity and formations rather than using a bottom-up approach to examine how group members perceive one another and their motivations for behaviour (Templeton et al., 2015). To fully understand the reasons for group clustering in the current research, the questionnaires could have included items measuring the extent to which students saw others as group members (for example items see (Doosje et al., 1995)). This would have allowed a unique combination of understanding both the physical and psychological factors contributing to the observed group behaviour.

It is worth noting that some limitations can affect the results. There was individual variability in each experimental condition that limits direct comparison, e.g., some students chose to individually enter or leave the room, and it was not possible to monitor some students since they were obscured from the camera by other people. In addition, the sample dimension was affected by external factors. In fact, unfortunately, during the research, the decision of the Dutch government to stop lectures again limited the number of the performed tests as well as the number of participants.

# 4 CONCLUSION

As some institutions and universities prepare to re-open or provide in-class activities, student adherence to the COVID-19 restrictions is a priority topic that requires close analysis. Proper physical distancing according to national and international health organisations can reduce individual exposure to COVID-19 and facilitate the reopening of buildings. This paper provides information on how students keep "safe" physical distances in common spaces of a university building and assess the extent to which students feel vulnerable to the COVID-19 exposure factors while moving in the building. An experimental-based approach was adopted in a case study application by focusing on movement of students along the corridors and while crossing through doorways.

Results could first be useful to support the development of behavioural models that describe individuals' COVID-19 exposure in universities. In addition, the results can provide interesting insights for the educational decision-makers regarding key risk areas in buildings (e.g., frequently touched surface areas, areas requiring close physical proximity, students' perceived vulnerability in different building areas).

From this perspective, results could be used by decision-makers to evaluate the required time for students to move inside the building (e.g., when changing between lessons) and to ensure sufficient physical capacity to maintain physical distancing (e.g., by phased timetabling to minimise occupancy of the corridors at any one time). Overall, results seem to suggest a need to encourage single lane formations in corridors to decrease the probability of students being in close physical proximity. However, we acknowledge that single lane formations in corridors could highly impact the use of the building and, in particular, the pedestrian flow through the building. Thus, since the work focuses on unidirectional flows in simplified straight corridor scenarios, further studies on other conditions (e.g., bidirectional flow, crossings) and other spatial constraints (e.g., multiple doors, different geometric shapes of the corridors) should be investigated to determine possible differences in how the students use the building. This data could additionally provide evidence on the effectiveness of one-way traffic paths in buildings in encouraging pedestrians to physically distance, as well as ideal layout shapes to support physical distancing strategies.

Further studies are also needed to enlarge the database, such as by investigating how types of buildings and how their uses impacts exposure to COVID-19 (i.e., those open to the public such as museums, transportation hubs and workplaces).

**ACKNOWLEDGMENTS**

The authors would like to thank all the students and staff who participated during the recordings and the distribution of the questionnaire.

**ETHICAL CONSIDERATIONS**

No personal identifiers of students or University staff have been shared with third parts and all results are presented in anonymized aggregate information and not at the individual level. All the information collected through the videos and the questionnaire are completely anonymous.

According to the Leiden University regulations in relation to the COVID-19, starting from 15 October it is compulsory for everyone in buildings to wear a face mask. Thus, students standing or walking, are required to wear a face mask. Recordings were taken only of people in movement (walking in the corridor, entering the room, sitting, exiting the room), with all the students wearing a face mask; this allowed investigators to avoid recording face and personal features that could be used to identify the participants.

The research received the confirmation of favourable opinion from the Faculteit Governance and Global Affairs Ethics Committee (Ethics application reference number: 2020-020-ISGA-Bartolucci)

Students were informed about the scope of the research; the questionnaire used for the data collection explicitly stated the scientific purpose of the study survey and the use of data; each responder provided verbal informed consent prior to participation and voluntarily decided whether to participate or not. Furthermore, students could withdraw from the research at any time.